\documentclass[a4paper,11pt]{article}

\usepackage{jheppub} 

\usepackage[T1]{fontenc} 
\usepackage{subcaption}
\usepackage{amsmath,amssymb,amsfonts}
\usepackage{bbold}
\usepackage{soul}
\usepackage[dvipsnames]{xcolor}
\usepackage{comment}
\usepackage{multirow}
\usepackage{tikz-feynman,contour}
\tikzfeynmanset{compat=1.1.0}

\usepackage{tikzsymbols}
\usepackage{array}
\usepackage{pifont} 

\title{\boldmath Leptogenesis from a $U(1)_D$ resonance}
\author[a]{E. J. Chun,}
\emailAdd{ejchun@kias.re.kr}
\affiliation[a]{School of School of Physics, Korea Institute for Advanced Study, Seoul 02455, Korea}
\author[b]{Arnab Dasgupta,}
\emailAdd{arnabdasgupta@protonmail.ch}
\affiliation[b]{ School of Liberal Arts, Seoul-Tech, Seoul 139-743, Korea}
\author[b,1]{Sin Kyu Kang \note{Corresponding author}}
\emailAdd{skkang@seoultech.ac.kr}

\abstract{
We propose a novel mechanism to realize leptogenesis  through the Breit-Wigner resonance of a dark $U(1)_D$ gauge boson $Z_D$, which mediates lepton number violating annihilations of dark matter (DM)
in the context of the scotogenic model with a $U(1)_D$. The processes occur out of equilibrium and the DM freezes out lately giving rise to the observed abundance.
The CP violation required for leptogenesis can be achieved by 
the interference between tree-level  t-channel scattering of DM and
the subsequent 1-loop mediated by $Z_D$, which arises
due to the unremovable imaginary part of either the $Z_D$ propagator coming from its self-energy correction or the 1-loop giving rise to the effective coupling of $Z_D\bar{\nu}\nu$.
}
\begin{document}

 \maketitle

\section{Introduction}

One of the great mysteries of our universe is the origin of the baryon asymmetry due to which net resultant baryons comprising almost 4\% of the entire energy budget of the Universe are generated. 
Long time ago, Sakharov found three ingredients to achieve the excess of baryon over anti-baryon through some processes,
which are
1) $B/L$ violation
2) $C$ and $CP$ violation and 3) a departure from thermal equilibrium. 
In addition to these conditions, the processes resulting in CP asymmetry
%
should abide by Nanopoulous-Weinberg theorem \cite{Nanopoulos:1979gx} and furthermore Adhikari-Rangarajan theorem \cite{Adhikari:2001yr}. 

There have been early attempts to generate baryon asymmetry from $1\rightarrow 2$ decay proposed in \cite{Dimopoulos:1978pw,Weinberg:1979bt} and also it can be dynamically generated via leptogenesis where lepton number, C and CP violating 2-body out-of-equilibrium decays of heavy Majorana neutrinos produce a primodial lepton asymmetry, which partially converted into the baryon asymmetry via $B+L$ violating sphaleron processes \cite{Fukugita:1986hr}.
However, leptogenesis through type I seesaw is successful for the mass scale of the heavy Majorana neutrinos larger than $\sim 10^{9}$ GeV \cite{Davidson:2002qv}, which is undesirable due to the hopeless possibility to probe it at collider experiments and a tension with naturalness of Higgs potential \cite{Vissani:1997ys, Clarke:2015gwa}.  
%

Motivated by those problems in vanilla leptogenesis, in this work, we propose a new promising possibility for leptogenesis realized naturally at low scale.
The major difficulty to achieve leptogenesis at low scale such as 1-10 TeV is that the condition for our of equilibrium decay in general requires very small coupling, which in turn generates lepton asymmetry insufficiently at low scale.
To remedy this, mass degeneracy of decaying particles and hierarchy of couplings that participate in the lepton number violating decay processes have been suggested leading to the realization of resonant leptogenesis\cite{Pilaftsis:1997jf}. But, they are rather unnatural to fit the right amount of the lepton asymmetry.
More natural solution for low scale leptogenesis 
would be generation of the lepton asymmetry through $2\rightarrow 2$ scatterings or three body decays \cite{Hambye:2001eu}. This is due to
the fact that the scattering cross section or decay rate constrained by the out of equilibrium condition is presented in terms of a quartic expression in two couplings
while the lepton asymmetry is in general only quadratic in a coupling.
Then, the lepton asymmetry can be naturally large enough with  sufficiently suppressed rates of the processes even at low scale.
In this work, we take into account the possibility of the lepton asymmetry generated through $2\rightarrow 2$ scatterings.
%

On the other hand, about 25$\%$ of the content of the Universe is constituted by the mysterious dark matter (DM).
While the nature of DM and the mechanism behind baryogenesis might be
uncorrelated to each other, it is tempting to construct models unifying both origins.
A plethora of attempts has been proposed in the recent years to explain the coincidence between baryon asymmetry and DM, $\Omega_{\rm DM} \approx 5 \Omega_{B}$, discarding simple numerical coincidence as an explanation to
the closeness of both abundances.
To incorporate low scale leptogenesis with DM, a plausible option is to generate lepton asymmetry and DM abundance simultaneously through  DM annihilations into a pair of leptons
\cite{Kumar:2013uca, Racker:2014uga, Dasgupta:2016odo, Borah:2018uci,Hugle:2018qbw}.
To achieve the goal with it, the processes occur out of equilibrium and the DM freezes out lately giving rise to the observed abundance.

In this scenario, we propose a new novel way of generating lepton asymmetry
through resonance of $Dark$ $U(1)_D$ gauge boson, $Z_D$, which mediates lepton number violating ($\Delta L=2$) DM annihilations.
In fact, those processes do not occur at tree level because of no lepton number violating neutral current mediated by $Z_D$.
However, as will be shown later, they are possible through chiral breaking effective vertex at 1-loop level.
To get CP violation required for leptogenesis, we take into account another DM annihilation through $t-$channel exchange of vector-like fermions.
Then, the interference of those two distinct amplitudes results in CP violation when there exists the unremovable imaginary part stemming from
either the $Z_D$ propagator coming from its self-energy correction or the effective vertex at 1-loop. This mechanism of generating the asymmetry from the imaginary part of the propagator has been pointed out by Dasgupta \emph{et. al.}\cite{Dasgupta:2019lha}.

To show how the new way of leptogenesis works, we adopt the framework of the scotogenic model where tiny Majorana neutrino masses naturally generated at 1-loop and there exist DM candidates.  
We extend the standard model(SM) gauge symmetry by introducing a $Dark$ $U(1)$ gauge symmetry. By spontaneously breaking the $Dark$ $U(1)$, the essential ingredient in scotogenic scenario, a $\mathbb{Z}_2$ symmetry,  naturally appears and tiny neutrino masses are generated at 1-loop. 
The neutral components of new scalar doublet will be responsible for DM
and their co-annihilations lead to the lepton asymmetry as explained above.
%
%
%
It is shown that such discovery and/or constraint can connect to the matter anti-matter asymmetry through leptogenesis.

This article is arranged as follows: first in section \ref{sec:model} we discuss about the model and its details regarding the mass spectrum of the scalars and the additional fermions, in section \ref{sec:lepto} we describe how the asymmetry can be generated through the resonance of the \emph{Dark} gauge $Z_D$ boson, in section \ref{sec:result} we present the result and then conclude,
and some useful formulae are presented in appendix.
\section{Model}
\label{sec:model}
For our purpose, we extend the SM gauge symmetry by introducing a $U(1)_D$.
To radiatively generate tiny neutrino masses and have natural DM candidates, we take the framework of the scotogenic model in which vector-like neutral fermions \footnote{ Note that we take into account vector-like neutral fermions instead of right-handed neutrinos, which is diiferent from the minmal scotogenic model.} and a new scalar doublet $\eta$ are introduced. We also introduce a singlet scalar field $\phi$ and a $SU(2)_L$ triplet scalar field $\Delta$ to break the $U(1)_D$ symmetry and
to naturally generate tiny mass splitting between neutral components of $\eta$, respectively.
 The complete contents of the matter fields in the model is given in table \ref{tab:model}.
 We note that the model is anomaly free.

\begin{table}[h!]
    \centering
    \begin{tabular}{|c|c|c|}
       \hline  
       & $SU(3)_c\times SU(2)_L \times U(1)_Y\times U(1)_D$ & F(fermion)/S(Scalar)\\ \hline
        $Q = \begin{pmatrix}u\\d\end{pmatrix}$ & $(3,2,1/6,0)$ & F \\
        $L = \begin{pmatrix}\nu\\e\end{pmatrix}$ & $(1,2,-1/2,0)$ & F \\
        $\bar{u}$ & $(\bar{3},1,-2/3,0)$ & F \\
        $\bar{d}$ & $(\bar{3},1,1/3,0)$ & F \\
        $\bar{e}$ & $(2,1,1,0)$ & F \\
        $H = \begin{pmatrix}h^0\\h^+\end{pmatrix}$ & $(1,2,-1/2,0)$ & S \\ \hline
        $\eta = \begin{pmatrix}\eta^-\\\eta^0\end{pmatrix}$ & $(1,2,-1/2,1)$ & S \\
        $\bar{\xi}$ & $(1,1,0,1)$ & F \\
        $\xi$ & $(1,1,0,-1)$ & F \\
        $\phi$ & $(1,1,0,2)$ & S \\
        $\Delta$ & $(1,3,-1,-2)$ & S \\ \hline
    \end{tabular}
    \caption{Summary of the matter fields. The upper half of the table corresponds to the SM and the lower half of the table to new particles. }
    \label{tab:model}
\end{table}
The Lagrangian for the entire scalar sector is given as 
\begin{align*}
\mathcal{L}_s &= \mu^2_H |H|^2 + \mu^2_\eta |\eta|^2 + \mu^2_\Delta |\Delta|^2 + \mu^2_\phi \phi^\dagger \phi + \lambda_H (H^\dagger H)^2 + \lambda_{H\eta} (H^\dagger \eta)(\eta^\dagger H) + \lambda^\prime_{H\eta} (H^\dagger H)(\eta^\dagger \eta) \nonumber \\
&+ \lambda_{H\phi}(H^\dagger H)\phi^* \phi + \lambda_\eta (\eta^\dagger \eta)^2 + \lambda_{\eta \phi}(\eta^\dagger \eta)\phi^* \phi + \lambda_\phi (\phi^* \phi)^2 + \lambda_\Delta {\rm Tr}[\Delta^\dagger \Delta]^2 + \lambda^\prime_\Delta {\rm Tr}[\Delta^\dagger \Delta \Delta \Delta^\dagger] \nonumber \\
&+ \mu_{\eta \Delta} \widetilde{\eta}^\dagger \Delta \eta + \lambda_{\eta \Delta} \eta^\dagger \eta {\rm Tr}[\Delta^\dagger \Delta] + \lambda^\prime_{\eta \Delta}\eta^\dagger \Delta^\dagger \Delta \eta + \lambda_{H\Delta} H^\dagger H {\rm Tr}[\Delta^\dagger \Delta] \nonumber \\
&+\lambda^\prime_{H\Delta} {\rm Tr}[H^\dagger\Delta^\dagger\Delta H] + \lambda_6 \widetilde{H}^\dagger\Delta H \phi.
\end{align*}
In this setup the $U(1)_D$ symmetry is broken to $\mathbb{Z}_2$ symmetry by the vacuum of $\phi$, which make the 
lightest neutral component of the doublet $\eta$ DM candidate.
 The scalar $\Delta$ also gets vacuum expectation value (VEV) along with $\phi$ breaking both $U(1)_D$ and $SU(2)_L$, however the VEV is taken to be very small as it is responsible for the mass splitting between two neutral components of the scale $\eta$.

\begin{figure}
    \centering
    \begin{tikzpicture}[/tikzfeynman/small]
    \begin{feynman}
    \vertex (i){$\nu_\alpha$};
    \vertex [right = 1.cm of i] (v1);
    \vertex [right = 1.cm of v1] (v2);
    \vertex [right = 1.cm of v2] (v3);
    \vertex [right = 1.cm of v3] (j){$\nu_\beta$};
    \vertex [blob,above = 1.cm of v2] (v4);
    \vertex [above = 0.4cm of v4] (k){$\langle \Delta^0 \rangle$};
    \vertex [below = 0.4cm of v2] (l){$\langle \phi \rangle$};
    \diagram*[small]{(i) -- [fermion](v1)--[fermion,edge label'=$\xi_i$](v2)--[anti fermion,edge label'=$\xi_i$](v3)--[anti fermion](j),
    (v1)--[scalar,quarter left,edge label=$\eta$](v4)--[scalar,quarter left,edge label=$\eta$](v3),
    (k)--[scalar](v4),(l)--[scalar](v2)};
    \end{feynman}
    \end{tikzpicture}
    \caption{One-loop diagram for neutrino mass.}
    \label{fig:numass}
\end{figure}
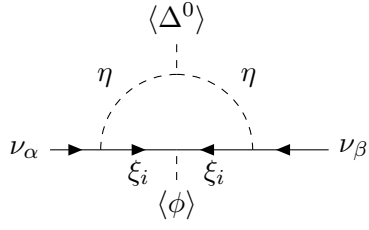

Solving the tadpole equation, one can get the following mass spectrum for the scalar sector in \{$H,\Delta^0,\phi$\} basis is given as

\begin{align*}
    m^2_s &= \begin{pmatrix}
    \lambda_H v^2 & 2(\lambda_{H\Delta} + \lambda^\prime_{H\Delta})v v_\Delta - \lambda_6 v v_\phi & (2\lambda_{H\phi}v_\phi - \lambda_6 v_\Delta)v \\
    2(\lambda_{H\Delta} + \lambda^\prime_{H\Delta})v v_\Delta - \lambda_6 v v_\phi & \frac{\lambda_6}{2v_\Delta}v^2v_\phi - 2 \lambda_\Delta v^2_\Delta & \frac{\lambda_6}{2}v^2 \\ 
    (2\lambda_{H\phi}v_\phi - \lambda_6 v_\Delta)v &  \frac{\lambda_6}{2}v^2 & \frac{\lambda_6}{2v_\phi}v^2v_\Delta + \lambda_\phi v^2_\phi
    \end{pmatrix},
\end{align*}
where $v$ is the vacuum of the SM Higgs, $v_{\phi}=\langle \phi^0 \rangle$ and
$v_{\Delta}=\langle\Delta^0 \rangle$,
and the pseudo-scalar is given as 
\begin{align}
    m^2_A &= \frac{\lambda_6}{2v_\Delta v_\phi}\left[4v^2_\Delta v^2_\phi + v^2(v^2_\Delta + v^2_\phi)\right].
\end{align}
The masses of the charged scalars are given as 
\begin{align}
    m^2_{\Delta^\pm} &= \frac{1}{2}\left(\lambda^\prime_{H\Delta} + \frac{\lambda_6}{v_\Delta}v_\phi\right)(v^2 + 2v^2_\Delta) \quad , \quad m^2_{\Delta^{\pm \pm}} = 4\lambda^\prime_{\Delta}v^2_\Delta + v^2\left[\lambda^\prime_{H\Delta} + \frac{1}{2}\frac{\lambda_6}{v_\Delta}v_\phi \right], \\
    m^2_{\eta^\pm} &= \mu^2_\eta +   \frac{1}{2}\lambda^{\prime}_{H \eta} v^2 + \lambda_{\eta \phi} v^2_\phi.
\end{align}
Now, for the limit $v_\phi/v_\Delta=\epsilon \gg 1$ the masses for the triplet become $m_{\Delta^{\pm\pm}}=m_{\Delta^{\pm}}=m_{\Delta_{R,I}^0}\equiv m_{\Delta}  \simeq \lambda_6/2v^2\epsilon$.
And finally the mass spectrum of the neutral components of $\eta$ are given as 
\begin{align}
    m^2_{\eta_R} &=  \mu^2_\eta + \frac{1}{2}(\lambda_{H\eta } +\lambda^\prime_{H\eta})v^2 + \sqrt{2}\mu_{\eta \Delta}v_{\Delta} + \lambda_{\eta \phi}v^2_\phi,  \\
     m^2_{\eta_I} &= \mu^2_\eta + \frac{1}{2}(\lambda_{H\eta } +\lambda^\prime_{H\eta})v^2 - \sqrt{2}\mu_{\eta \Delta}v_{\Delta} + \lambda_{\eta \phi}v^2_\phi. 
\end{align}
The mass splitting between the real and imaginary parts is controlled by $\mu_{\eta \Delta}$.
Now,  the Lagrangian for the fermionic sector is given as 
\begin{align}
    \mathcal{L} &= m_{\xi} \xi \bar{\xi} + Y_u  \overline{Q} H \bar{u} + Y_d \widetilde{H}^\dagger Q \bar{d} + Y_l H^\dagger L \bar{e} + Y_\nu \widetilde{\eta}^\dagger L \xi \nonumber \\
    &+ Y_{L} \phi^* \xi \xi + Y_{R} \phi \bar{\xi}\bar{\xi} + h.c .
\end{align}
All the quarks and leptons acquire the masses through standard higgs mechanism. 
From the above equation the mass matrix of the vector-like fermions $\xi,\overline{\xi}$ comes out to be 
\begin{align}
    M &= \begin{pmatrix} Y_L v_{\phi} & m_\xi \\ m_\xi & Y_R v_{\phi}\end{pmatrix},
\end{align}
in the interaction basis of $\{\xi,\overline{\xi}\}$. Assuming the $m_\xi,Y_L$ and $Y_R$ to be diagonal, the mixing angle $\theta$ for each generation of $\xi$'s that diagonalizes this mass matrix is given as 
\begin{align}
    \tan{2\theta_i} &= \frac{2m_{\xi_i}}{(Y_{L_i} - Y_{R_i})v_{\phi}}.
    \label{eq:mix}
\end{align}
 And hence the  vector-like fermions masses are given as 
\begin{align}
    M_{i^{\pm}} &= (Y_{L_i} +Y_{R_i})\frac{v_\phi}{\sqrt{2}} \pm \frac{1}{2}\sqrt{(Y_{L_i} - Y_{R_i})^2v^2_\phi + 2m^2_{\xi_i}},
\end{align}
and finally the neutrinos will acquire the mass at loop shown in fig \ref{fig:numass} and is given as 
\begin{align}
    (m_{\nu})_{\alpha \beta} &= \sum_i Y_{\nu_{i \alpha}}Y_{\nu_{i \beta}}\left(\cos^2{\theta_i}\frac{M_{i^+}}{32\pi^2}\left[\frac{m^2_{\eta_R}}{m^2_{\eta_R}-M^{2}_{i^+}}\ln\left(\frac{m^2_{\eta_R}}{M^{2}_{i^+}}\right) - \frac{m^2_{\eta_I}}{m^2_{\eta_I}-M^{2}_{i^+}}\ln\left(\frac{m^2_{\eta_I}}{M^{+2}_{i}}\right) \right] \right. \nonumber \\ 
    &+ \left. \sin^2{\theta_i}\frac{M_{i^-}}{32\pi^2}\left[\frac{m^2_{\eta_R}}{m^2_{\eta_R}-M^{2}_{i^{-}}}\ln\left(\frac{m^2_{\eta_R}}{M^{2}_{i^-}}\right) - \frac{m^2_{\eta_I}}{m^2_{\eta_I}-M^{2}_{i^-}}\ln\left(\frac{m^2_{\eta_I}}{M^{2}_{i^-}}\right) \right]\right), \nonumber \\
    (m_{\nu})_{\alpha \beta} &= \left(Y_\nu^T\Lambda Y_\nu\right)_{\alpha \beta}, \nonumber \\
    \Lambda_i &= \left(\cos^2{\theta_i}\frac{M_{i^+}}{32\pi^2}\left[\frac{m^2_{\eta_R}}{m^2_{\eta_R}-M^{2}_{i^+}}\ln\left(\frac{m^2_{\eta_R}}{M^{2}_{i^+}}\right) - \frac{m^2_{\eta_I}}{m^2_{\eta_I}-M^{2}_{i^+}}\ln\left(\frac{m^2_{\eta_I}}{M^{2}_{i^+}}\right) \right] \right. \nonumber \\ 
    &+ \left. \sin^2{\theta_i}\frac{M_{i^-}}{32\pi^2}\left[\frac{m^2_{\eta_R}}{m^2_{\eta_R}-M^{2}_{i^-}}\ln\left(\frac{m^2_{\eta_R}}{M^{2}_{i^-}}\right) - \frac{m^2_{\eta_I}}{m^2_{\eta_I}-M^{2}_{i^-}}\ln\left(\frac{m^2_{\eta_I}}{M^{2}_{i^-}}\right) \right]\right).
\end{align}
One may recall that $\Lambda_i = 0$ if $m_{\eta_R}=m_{\eta_I}$ ($v_\Delta = 0$) or $M_{i^+}=-M_{i^-}$($v_\phi = 0$ ) for which a conserve $U(1)_L$ can be defined.
For the sake of numerical analysis satisfying the observed neutrino oscillation data, 
we take the Casas-Ibarra parameterization of the Yukawa ($Y_\nu$) as follows;
\begin{align}
Y_{\nu_{i\alpha}} &= \left(\sqrt{\Lambda}^{-1}R\sqrt{m^{diag}_\nu}U^\dagger_{\rm PMNS}\right)_{i\alpha}.
    \label{eq:CI}
\end{align}
Here, $U_{\rm PMNS}$ is the so-called PMNS neutrino mixing matrix, $R$ is general complex orthogonal matrix and $m^{diag}_\nu = {\rm Diag}(m_1,m_2,m_3)$. In our case the general complex matrix $R$ can be parameterized by three complex parameters of $\theta_{\alpha \beta} = \theta^R_{\alpha \beta} + i \theta^I_{\alpha \beta} \in [0,2\pi]$. In general, the orthogonal matrix $R$ for $n$ flavors can be $^nC_2$ number of rotation matrices of type:
\begin{align}
R_{\alpha \beta} &= \begin{pmatrix} \cos{(\theta^R_{\alpha \beta} + i \theta^I_{\alpha \beta})} & \cdots & \sin{(\theta^R_{\alpha \beta} + i \theta^I_{\alpha \beta})} \\ 
\vdots & \ddots & \vdots \\
-\sin{(\theta^R_{\alpha \beta} + i \theta^I_{\alpha \beta})} & \cdots & \cos{(\theta^R_{\alpha \beta} + i \theta^I_{\alpha \beta})}\end{pmatrix}
\end{align}
In this scenario, since we take the scalar doublet $\eta$ as the dark matter, all the $\xi$'s are decayed and do not take part in the Boltzmann evolution.

\section{Leptogenesis through Resonance of $U(1)_D$ gauge boson}
\label{sec:lepto}
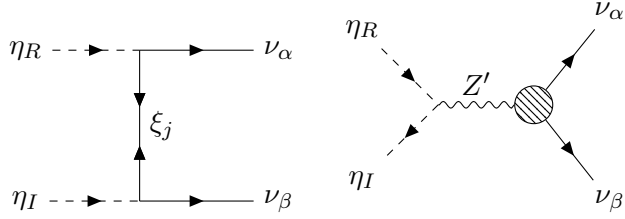
\begin{figure}
    \centering
    \begin{tabular}{lr}
    \begin{tikzpicture}[/tikzfeynman/small]
    \begin{feynman}
    \vertex (i){$\eta_R$};
    \vertex[below = 2.cm of i](j){$\eta_I$};
    \vertex[right = 1.5cm of i](v1);
    \vertex[right = 1.5cm of v1](k){$\nu_\alpha$};
    \vertex[right = 1.5cm of j](v2);
    \vertex[right = 1.5cm of v2](l){$\nu_\beta$};
    \diagram*[small]{(i)--[charged scalar](v1),(v1)--[majorana,edge label=$\xi_j$](v2),(v2)--[fermion](l),
    (v1)--[fermion](k),(j)--[charged scalar](v2)};
    \end{feynman}   
    \end{tikzpicture}&
    \begin{tikzpicture}[/tikzfeynman/small]
    \begin{feynman}
    \vertex (i){$\eta_R$};
    \vertex[below = 2.cm of i](j){$\eta_I$};
    \vertex[below = 1.cm of i](v1);
    \vertex[right = 1.cm of v1](v2);
    \vertex[right = 1.cm of v2,blob](v3){};
    \vertex[right = 1.cm of v3](v4);
    \vertex[above = 1.cm of v4](k){$\nu_\alpha$};
    \vertex[below = 1.cm of v4](l){$\nu_\beta$};
    \diagram*[small]{(i)--[charged scalar](v2) --[charged scalar](j),(v2)--[boson,edge label=$Z^\prime$](v3)--[fermion](k),(v3)--[fermion](l)};
    \end{feynman}   
    \end{tikzpicture}
    \end{tabular}
    \caption{Two distinct amplitudes whose interference gives rise to lepton asymmetry.}
    \label{fig:amplepto}
\end{figure}

In this section we discuss the mechanism of generating the lepton asymmetry through the Breit-Wigner resonance of the $U(1)_D$ gauge boson which mediates lepton number violating ($\Delta L=2$) neutral current. 
In appendix \ref{sec:app1}, we present how lepton number violating neutral currents
are generated at 1-loop, which give rise to effective vertex as explicitly presented in
eq.(\ref{effectiveC}).
Thanks to the effective coupling given in eq.(\ref{effectiveC}), lepton number violating
$2\rightarrow 2$ processes mediated by $Z^{\prime}_D$ can arise as shown in Fig. \ref{fig:amplepto} in which the blob represents the effective coupling.
In addition, lepton number is violated through the exchange of $\xi_i$ as shown in Fig. \ref{fig:amplepto}.
Then, CP violating lepton asymmetry can be obtained from the interference of the amplitudes of
both lepton number violation processes.
The general expression for the lepton asymmetry arising from the difference of the amplitude square of the particle and anti-particle is given as 
\begin{align}
    \delta \equiv |\mathcal{M(\eta \eta \rightarrow \nu \nu)}|^2-|\mathcal{M(\eta \eta \rightarrow \bar{\nu} \bar{\nu})}|^2= 4\Im[C^*_0C_1]\Im[\mathcal{A}^*_0\mathcal{A}_1], \label{delta}
\end{align}
where the first part of the asymmetry (i.e $\mathcal{C}$'s) arises only from the multiplication of the couplings in two amplitudes and the second one (i.e $\mathcal{A}$'s) comes from the pure amplitude except for the couplings. 
The effective amplitudes are given as 
\begin{align}
    i\mathcal{M}_0 =C_0\mathcal{A}_0 &= i\sum_i Y_{i\alpha}Y_{i\beta} (x^\dagger_\alpha x^\dagger_\beta)\left( \cos^2{\theta_i}M_{i^+}\left[\frac{1}{t-M^2_{i^+}} + \frac{1}{u-M^2_{i^+}}\right]\right. \nonumber \\
    &+ \left. \sin^2{\theta_i}M_{i^-}\left[\frac{1}{t-M^2_{i^-}} + \frac{1}{u-M^2_{i^-}}\right]\right), \nonumber \\
     i\mathcal{M}_1 =C_1\mathcal{A}_1 &= ig^2\sum_i Y_{i\alpha}Y_{i\beta}f_i \frac{x^\dagger\bar{\sigma}_{\mu \nu}x^\dagger_\beta(p_\alpha-p_\beta)(p_R-p_I)_\nu}{(s-m^2_{Z_D} + i M_{Z_D}\Gamma_{Z_D})}.
\end{align}
The above amplitudes are written in terms of two-spinor notation where $x^\dagger$'s are the \emph{commutating} two-component spinor wave function as explained in \cite{Dreiner:2008tw}. 

The loop function $f_i$ is given in eq.(\ref{effectiveC}). In order to understand the dependence of the above asymmetry with respect to the temperature it is convenient to separate $\delta$ in eq.(\ref{delta}) into two contributions as follows: 
\begin{align}
    \delta_R &= \sum_{\alpha,\beta} \frac{\Im[Y^*_{j\alpha}Y^*_{j\beta}Y_{i\alpha}Y_{i\beta}]g^2\Re[f_i]m_{Z_D}\Gamma_{Z_D}}{((s-m^2_{Z_D})^2 + m^2_{Z_D}\Gamma^2_{Z_D})}(m^2_{\eta_R}-m^2_{\eta_I})\frac{(s+m^2_{Z_D})}{m^2_{Z_D}}\nonumber \\
    &\times\left( \cos^2{\theta_i}M_{j^+}\left[\frac{1}{t-M^2_{j^+}} + \frac{1}{u-M^2_{j^+}}\right] + \sin^2{\theta_i}M_{j^-}\left[\frac{1}{t-M^2_{j^-}} + \frac{1}{u-M^2_{j^-}}\right]\right), \nonumber \\
    \delta_I &= \sum_{\alpha,\beta} \frac{\Im[Y^*_{j\alpha}Y^*_{j\beta}Y_{i\alpha}Y_{i\beta}]g^2\Im[f_i](s-m^2_{Z_D})}{((s-m^2_{Z_D})^2 + m^2_{Z_D}\Gamma^2_{Z_D})}(m^2_{\eta_R}-m^2_{\eta_I})\frac{(s+m^2_{Z_D})}{m^2_{Z_D}}\nonumber \\
    &\times\left( \cos^2{\theta_i}M_{j^+}\left[\frac{1}{t-M^2_{j^+}} + \frac{1}{u-M^2_{j^+}}\right] + \sin^2{\theta_i}M_{j^-}\left[\frac{1}{t-M^2_{j^-}} + \frac{1}{u-M^2_{j^-}}\right]\right). 
    \label{deltas}
\end{align}
We observe that $\delta_R$ ($\delta_I$) is proportional to the real (imaginary) part of chirality violating 
vertex of the $Z_D$.
Here, we note that $\delta_{R,I}= 0$ for 
$m_{\eta_R}=m_{\eta_I}$(or $M_{i^+}=M_{i^-}$) restoring $U(1)_D$ conservation. 
Thus, the asymmetry is suppressed for $m_{\eta_R} \approx m_{\eta_I}$, which 
eliminates the effect of the $Z_D$ resonance. 
Note also that the asymmetry $\delta_R \propto \Gamma_D$ comes from the 1-loop diagram involving the $Z_D\rightarrow \eta \eta$ or $\xi \xi$ vertices which are independent of the tree-level diagram.

Then, the corresponding thermally averaged cross-sections are given as 
\begin{align}
\gamma^\delta_R &= \frac{T}{8\pi^4}\int^\infty_{s_{in}}\int^1_{-1}\delta_R \frac{p_{in}p_{out}}{\sqrt{s}}K_1(\sqrt{s}/T)~ d(\cos{\theta})~ ds,\label{eq:delR} \\
\gamma^\delta_I &= \frac{T}{8\pi^4}\int^\infty_{s_{in}}\int^1_{-1}\delta_I \frac{p_{in}p_{out}}{\sqrt{s}}K_1(\sqrt{s}/T)~ d(\cos{\theta})~ds,\label{eq:delI}
\end{align}
where $K_1$ is the modified Bessel function of the second kind.

Now, in order to get the lepton asymmetry along with the relic abundance we solve the coupled Boltzmann equation given as
\begin{align}
    \frac{dX_{DM}}{dz} &= \frac{1}{zsH(M_{\eta_R})}\left(\frac{X^2}{(X^{eq}_{DM})^2}-1\right)
    \gamma^{eq}_{scatt}(DM DM \rightarrow SM SM), \nonumber \\
    \frac{dX_L}{dz} &= \frac{1}{zsH(M_{\eta_R})} \left[\left(\frac{X_{\eta_R}X_{\eta_I}}{X^{eq}_{\eta_R}X^{eq}_{\eta_R}}-1\right)(\gamma^\delta_R - \gamma^\delta_I) \right. \nonumber \\
    &-  \frac{X_L}{X^{eq}_l}\left(2\gamma^{eq}_{scatt}(\eta \eta \rightarrow L L) + \gamma^{eq}_{scatt}(\eta \xi \rightarrow L SM) + \gamma^{eq}_{scatt}(\eta L \rightarrow \eta \bar{L}) \right. \nonumber \\
    &+ \left. \left.\gamma^{eq}_{scatt}(\eta L \rightarrow \xi SM) + \gamma_{D}(\xi \rightarrow \eta L) + \gamma^{eq}_{scatt}(\xi L\rightarrow \eta SM)\right)\right],
    \label{BZE}
\end{align}
where $DM \in \{\eta_R,\eta_I,\eta_\pm\}$ and $X_i = n_i/s$ are the comoving number density in which $n_i$ is the number density and $s=g_*2\pi^2/45 T^3$ is the entropy density with $g_*$ being the number of relativistic degree of freedom.
The first equation in eq.(\ref{BZE}) is for the evolution of number density of dark sector and the second one is for that of lepton number asymmetry. 
In the second equation, the first line of right-handed side corresponds to
the generation of lepton asymmetry, whereas the other two lines to the washout.
\section{Results and Discussion}
\label{sec:result}
In this section we present numerical results for the evolution of the lepton number asymmetry along with the relic abundance of DM. 
The important point to be noticed before presenting our results is that the contribution coming from the 
resonance of the $Z_D$ to the asymmetry in Eq.(\ref{delta}) is always dominant over the one coming from the imaginary part of the loop function. This due to the reason that near the resonance point $s\sim m_{Z_D}$ the $\delta_I$ in Eq.(\ref{deltas}) goes to zero and thus the required baryon asymmetry can be achieved via $\delta_R$. 
This is a novel way to generate the asymmetry for successful baryogenesis.
But, to realize this mechanism through the resonance of the unstable neutral gauge boson, there should exist
another distinct amplitude which is interfered with the amplitude mediated by the neutral gauge boson. In our case it is  through the $t-$channel processes mediated by the vector-like fermion as shown in Fig. \ref{fig:amplepto}.

\begin{figure}[!h]
    \centering
    \begin{tabular}{lcr}
    \includegraphics[width=0.5\textwidth]{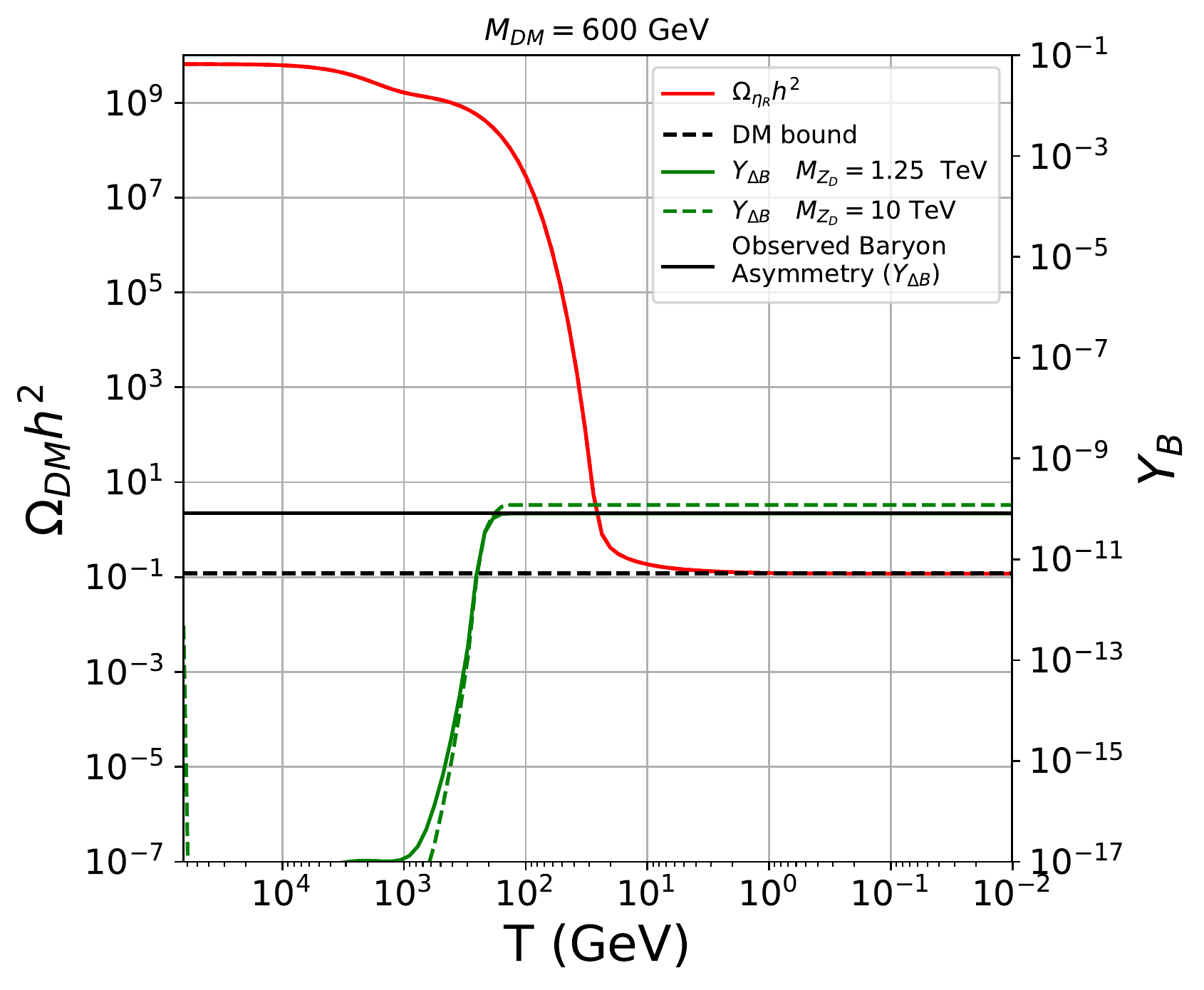}&
    \includegraphics[width=0.5\textwidth]{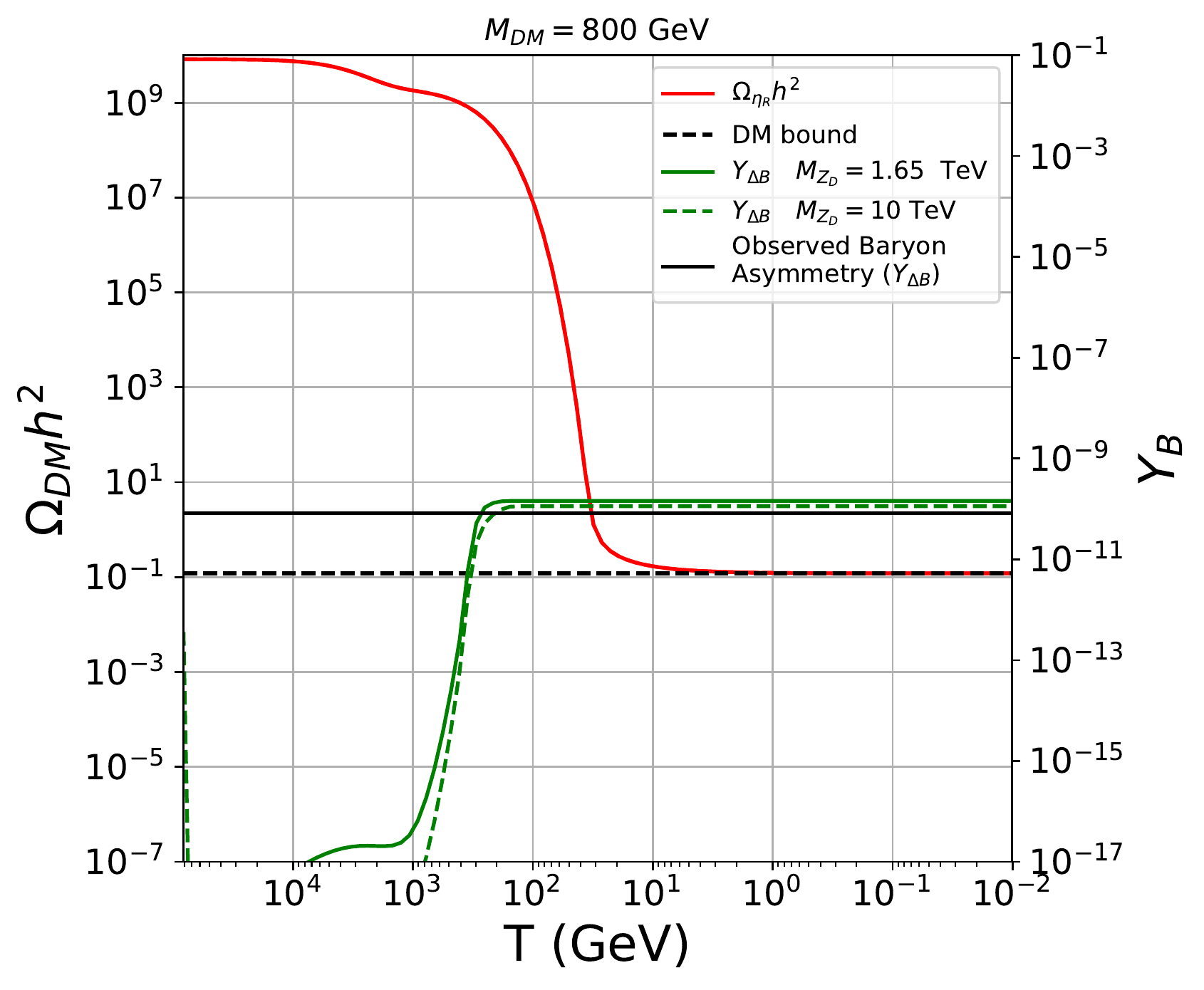}     
    \end{tabular}
    \caption{Number densities as a function of $T$. Red solid, green (solid and dashed) lines correspond to relic density and baryon number asymmetry, respectively. For the plots of baryon asymmetry, we take $M_{Z_D}$ to be 1.25 (green solid), 10 (green dashed) TeV  for $M_{DM} = 600$ GeV in the left panel, and 1.65 (green solid), 10 (green dashed) TeV as input for $M_{DM} = 800$ GeV in the right panel.}
    \label{fig:lepto}
\end{figure}

For the numerical analysis, we take the central values of neutrino oscillation data as input \cite{Esteban:2018azc}. The values of model parameters we take as a benchmark are presented in Table \ref{tab:BP2}. One may notice that we have taken $\lambda_{H\eta} = - \lambda^\prime_{H\eta}$ for all the Bench mark points, this is to make $\eta_R$ and $\eta_I$ trivially small with respect to $\eta^\pm$.
The plots in Fig. \ref{fig:lepto} show how the relic density of DM (red line) and
baryon number asymmetry (green lines) evolve along with temperature.
Their experimental results are given by
\begin{align}
\Omega_{\text{DM}} h^2= 0.120 \pm 0.001, \\
Y_{\Delta B} = 8.237\times 10^{-11}
\label{eq:PLANCK}
\end{align}
at 68\% CL~\cite{Aghanim:2018eyx}.
The equilibrium number density of DM tracks the red dashed line.
The left(right) panel corresponds to $M_{\text{DM}}=600 (800)$ GeV.
The solid green lines represent the prediction of  the baryon number asymmetry  for $m_{Z_D}$=
1.25 TeV (left panel) and 1.65 TeV (right panel), respectively, which are nearly $m_{Z_D}\sim 2M_{DM}$.
Note that our mechanism also works for $m_{Z_D} > 2M_{DM}$.
The plots show that the DM freezes out at $T_{f_D}\sim 30$ GeV, whereas
the baryon asymmetry freezes out at $T_{f_B} \sim 200-300$ GeV. 
One may notice that although the integrand in eq. \eqref{eq:delR} has a peak near the resonance (i.e $\sqrt{s}=m_{Z_D}$), $\gamma^{\delta}_R$ can be suppressed due to a Boltzmann 
suppression coming from the modified Bessel function ($K_1(m_{Z_D}/T_{f_B})$). When $T_{f_B}\simeq 200$ GeV the suppression is least for the case of $m_{Z_D}\simeq 2 m_{DM}$, 
for which the correct asymmetry is achieved for rather small values of $\theta^{R,I}_{ij}$.
On the other hand, for $m_{Z_D}> 2m_{DM}$, the right amount of the asymmetry can be obtained
only when $\theta^{R,I}_{ij}$ is large.
The green-dashed lines depicted in Fig. \ref{fig:lepto} correspond to this case, for which we take
$m_{Z_D}=10$ TeV. As presented in Table 2, the required values of $\theta_{ij}^{R(I)}$ to achieve right amount of baryon asymmetry in that case are 40 times larger than those for the cases corresponding to the solid green lines.
This is due to the fact that for larger $m_{Z_D}$  the resonance occurs at higher $\sqrt{s}$ value which makes the asymmetry 
suppressed by the Boltzmann factor mentioned above, so
taking larger values of $\theta^{R,I}_{ij}$ for fixed
other inputs  can help to compensate this Boltzmann suppression.
As can be seen from Fig. \ref{fig:lepto},  right amount of baryon asymmetry and relic density of DM can be simultaneously achieved in our scenario


\begin{table}[!b]
    \centering
    \begin{tabular}{|c|c|c|c|c|} \hline \hline
    &\multicolumn{2}{|c|}{BP1}& \multicolumn{2}{|c|}{BP2} \\\hline
    $v_\Delta$ & \multicolumn{4}{|c|}{1  GeV} \\ \hline
    $\mu_\eta$  & \multicolumn{2}{|c|}{600 GeV} & \multicolumn{2}{|c|}{800 GeV} \\ \hline
    $m_{N_i(=1,2,3)}$ & \multicolumn{2}{|c|}{6 TeV} & \multicolumn{2}{|c|}{8 TeV}  \\ \hline
    $m_{\eta^0_I}$ & 601 GeV & 678 GeV & 802 GeV & 860 GeV\\ \hline
    $\Delta m_{\eta^0}$ & 7.06 MeV & 6.25 MeV &  5.3 MeV & 4.93 MeV \\ \hline
    $m_{\eta^\pm}$ & 606 GeV & 685 GeV & 808 GeV & 868 GeV\\ \hline
    $ m_{\phi}$ & 3.95 TeV & 31.62 TeV & $5.22$ TeV & 31.62 TeV\\ \hline
    $m_{\Delta}$ & 603 GeV & 1.74 TeV & 707 GeV & 1.74 TeV \\ \hline
    $\lambda_H$ & \multicolumn{4}{|c|}{0.253} \\ \hline
    $\lambda_{H\eta}$ & 0.15 & 0.29  & 0.34 & 0.43 \\ \hline
    $\lambda^\prime_{H\eta}$ & -0.15 & -0.29  & -0.34 & -0.43\\ \hline
    $\theta^{R}_{13}=\theta^{I}_{13}$ & $\pi/800$& $\pi/20$ & $\pi/800$ & $\pi/20$\\ \hline
    $g_{D}$ & \multicolumn{4}{|c|}{0.05} \\ \hline
    $\Gamma_{Z_D}$ & $2.04\times 10^{-3}$ GeV & 1.96 TeV & 0.0454 GeV & 1.96 TeV\\ \hline
    $M_{Z_D}$ & 1.25 TeV & 10 TeV  & 1.65 TeV & 10 TeV \\ \hline
    $\mu_{\eta \Delta}$ & \multicolumn{4}{|c|}{3 GeV} \\ \hline \hline
    \end{tabular}
    \caption{Input values of the parameters for two benchmark points. Here $m_{\Delta}=m_{\Delta^{\pm\pm}}=m_{\Delta^{\pm}}=m_{\Delta_{R,I}^0}$. The final states of $Z_D$ decay into the scalar triplet ($\Delta^*, \Delta $) denoting $(\Delta^0_R,\Delta^0_I),(\Delta^{++},\Delta^{--}),(\Delta^+,\Delta^-)$ and to the scalar double fields $(\eta^*,\eta)$ denote $(\eta^0_R,\eta^0_I),(\eta^+,\eta^-)$.
    }
    \label{tab:BP2}
\end{table}

\section{Conclusion}
We have proposed a novel mechanism to generate baryon asymmetry  through the Breit-Wigner resonance of a dark $U(1)_D$ gauge boson $Z_D$, which mediates lepton number violating annihilations of dark matter (DM)
in the context of the scotogenic model with a $U(1)_D$. 
The origin of  CP asymmetry required for leptogenesis is the interference between tree-level  t-channel scattering of DM and
the subsequent 1-loop mediated by $Z_D$, which arises
due to the unremovable imaginary part of either the $Z_D$ propagator coming from its self-energy correction or the 1-loop vertex giving rise to the effective coupling of $Z_D\bar{\nu}\nu$. The former is always dominant over the latter thanks to the occurrence of resonance
of the gauge boson $Z_D$.
The processes occur out of equilibrium and the DM freezes out lately giving rise to the observed abundance.
We could show that  right amount of baryon asymmetry and relic density of DM can be simultaneously achieved in our scenario.

\appendix
\section{Details of the effective $Z_D$ coupling}
\label{sec:app1}

\begin{figure}
    \centering
     \begin{tikzpicture}[/tikzfeynman/small]
     \begin{feynman}
     \vertex (i){$Z_D$};
     \vertex[right = 1.5cm of i,blob](v1){};
     \vertex[right = 1.cm of v1](v2);
     \vertex[above = 1.cm of v2](k){$\nu_\alpha$};
     \vertex[below = 1.cm of v2](l){$\nu_\beta$};
     \diagram*[small]{(i) --[boson](v1)--[fermion](k),(v1)--[fermion](l)};
     \end{feynman}   
     \end{tikzpicture}
    \caption{Diagram for the effective vertex of $Z_D\nu \nu$.}
    \label{fig:eff_ZD}
\end{figure}
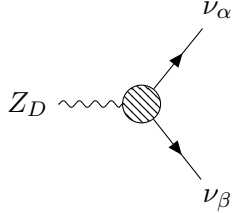

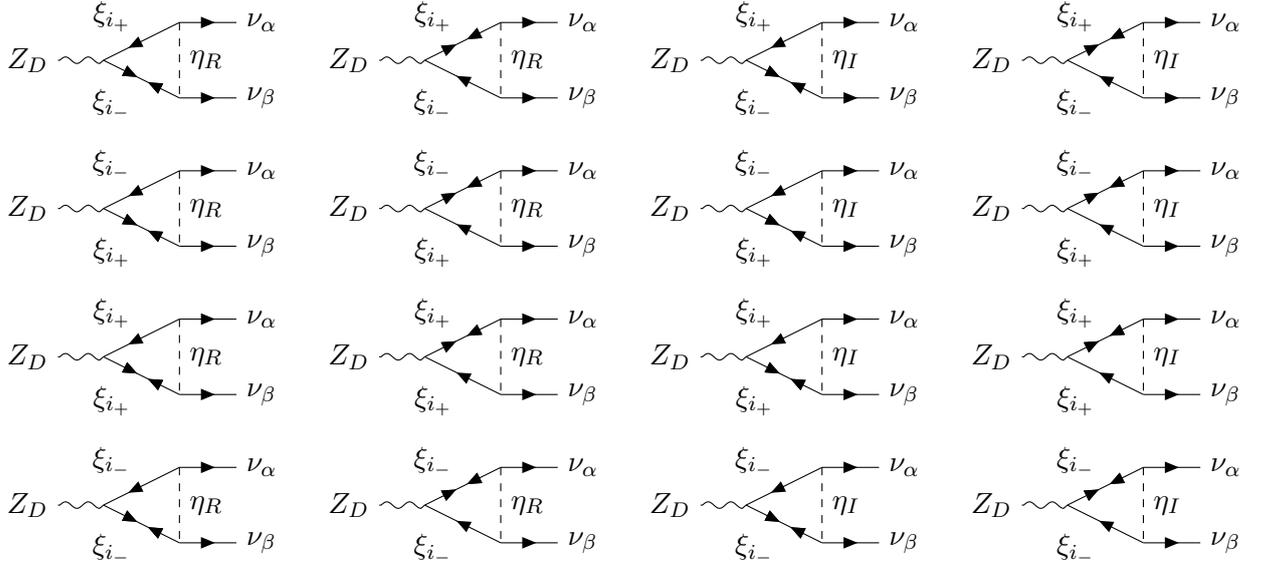
\begin{figure}[!h]
    \centering
    \begin{tabular}{lccr}
        \begin{tikzpicture}[/tikzfeynman/small]
     \begin{feynman}
     \vertex (i){$Z_D$};
     \vertex[right = 1.cm of i](v1);
     \vertex[right = 1.cm of v1](v4);
     \vertex[above = 0.5cm of v4](v2);
     \vertex[below = 0.5cm of v4](v3);
     \vertex[right = .75cm of v2](j){$\nu_{\alpha}$};
     \vertex[right = .75cm of v3](k){$\nu_{\beta}$};
     \diagram*[small]{(i) --[boson](v1),(v3)--[majorana,edge label = $\xi_{i_-}$](v1)--[anti fermion,edge label = $\xi_{i_+}$](v2),(v2)--[fermion](j),(v3)--[fermion](k),(v2)--[scalar,edge label=$\eta_R$](v3)};
     \end{feynman}   
     \end{tikzpicture} & 
     \begin{tikzpicture}[/tikzfeynman/small]
     \begin{feynman}
     \vertex (i){$Z_D$};
     \vertex[right = 1.cm of i](v1);
     \vertex[right = 1.cm of v1](v4);
     \vertex[above = 0.5cm of v4](v2);
     \vertex[below = 0.5cm of v4](v3);
     \vertex[right = .75cm of v2](j){$\nu_{\alpha}$};
     \vertex[right = .75cm of v3](k){$\nu_{\beta}$};
     \diagram*[small]{(i) --[boson](v1),(v3)--[fermion,edge label = $\xi_{i_-}$](v1)--[majorana,edge label = $\xi_{i_+}$](v2),(v2)--[fermion](j),(v3)--[fermion](k),(v2)--[scalar,edge label=$\eta_R$](v3)};
     \end{feynman}   
     \end{tikzpicture} &
    \begin{tikzpicture}[/tikzfeynman/small]
     \begin{feynman}
     \vertex (i){$Z_D$};
     \vertex[right = 1.cm of i](v1);
     \vertex[right = 1.cm of v1](v4);
     \vertex[above = 0.5cm of v4](v2);
     \vertex[below = 0.5cm of v4](v3);
     \vertex[right = .75cm of v2](j){$\nu_{\alpha}$};
     \vertex[right = .75cm of v3](k){$\nu_{\beta}$};
     \diagram*[small]{(i) --[boson](v1),(v3)--[majorana,edge label = $\xi_{i_-}$](v1)--[anti fermion,edge label = $\xi_{i_+}$](v2),(v2)--[fermion](j),(v3)--[fermion](k),(v2)--[scalar,edge label=$\eta_I$](v3)};
     \end{feynman}   
     \end{tikzpicture} &
     \begin{tikzpicture}[/tikzfeynman/small]
     \begin{feynman}
     \vertex (i){$Z_D$};
     \vertex[right = 1.cm of i](v1);
     \vertex[right = 1.cm of v1](v4);
     \vertex[above = 0.5cm of v4](v2);
     \vertex[below = 0.5cm of v4](v3);
     \vertex[right = .75cm of v2](j){$\nu_{\alpha}$};
     \vertex[right = .75cm of v3](k){$\nu_{\beta}$};
     \diagram*[small]{(i) --[boson](v1),(v3)--[fermion,edge label = $\xi_{i_-}$](v1)--[majorana,edge label = $\xi_{i_+}$](v2),(v2)--[fermion](j),(v3)--[fermion](k),(v2)--[scalar,edge label=$\eta_I$](v3)};
     \end{feynman}   
     \end{tikzpicture}\\
            \begin{tikzpicture}[/tikzfeynman/small]
     \begin{feynman}
     \vertex (i){$Z_D$};
     \vertex[right = 1.cm of i](v1);
     \vertex[right = 1.cm of v1](v4);
     \vertex[above = 0.5cm of v4](v2);
     \vertex[below = 0.5cm of v4](v3);
     \vertex[right = .75cm of v2](j){$\nu_{\alpha}$};
     \vertex[right = .75cm of v3](k){$\nu_{\beta}$};
     \diagram*[small]{(i) --[boson](v1),(v3)--[majorana,edge label = $\xi_{i_+}$](v1)--[anti fermion,edge label = $\xi_{i_-}$](v2),(v2)--[fermion](j),(v3)--[fermion](k),(v2)--[scalar,edge label=$\eta_R$](v3)};
     \end{feynman}   
     \end{tikzpicture} & 
     \begin{tikzpicture}[/tikzfeynman/small]
     \begin{feynman}
     \vertex (i){$Z_D$};
     \vertex[right = 1.cm of i](v1);
     \vertex[right = 1.cm of v1](v4);
     \vertex[above = 0.5cm of v4](v2);
     \vertex[below = 0.5cm of v4](v3);
     \vertex[right = .75cm of v2](j){$\nu_{\alpha}$};
     \vertex[right = .75cm of v3](k){$\nu_{\beta}$};
     \diagram*[small]{(i) --[boson](v1),(v3)--[fermion,edge label = $\xi_{i_+}$](v1)--[majorana,edge label = $\xi_{i_-}$](v2),(v2)--[fermion](j),(v3)--[fermion](k),(v2)--[scalar,edge label=$\eta_R$](v3)};
     \end{feynman}   
     \end{tikzpicture} &
    \begin{tikzpicture}[/tikzfeynman/small]
     \begin{feynman}
     \vertex (i){$Z_D$};
     \vertex[right = 1.cm of i](v1);
     \vertex[right = 1.cm of v1](v4);
     \vertex[above = 0.5cm of v4](v2);
     \vertex[below = 0.5cm of v4](v3);
     \vertex[right = .75cm of v2](j){$\nu_{\alpha}$};
     \vertex[right = .75cm of v3](k){$\nu_{\beta}$};
     \diagram*[small]{(i) --[boson](v1),(v3)--[majorana,edge label = $\xi_{i_+}$](v1)--[anti fermion,edge label = $\xi_{i_-}$](v2),(v2)--[fermion](j),(v3)--[fermion](k),(v2)--[scalar,edge label=$\eta_I$](v3)};
     \end{feynman}   
     \end{tikzpicture} &
     \begin{tikzpicture}[/tikzfeynman/small]
     \begin{feynman}
     \vertex (i){$Z_D$};
     \vertex[right = 1.cm of i](v1);
     \vertex[right = 1.cm of v1](v4);
     \vertex[above = 0.5cm of v4](v2);
     \vertex[below = 0.5cm of v4](v3);
     \vertex[right = .75cm of v2](j){$\nu_{\alpha}$};
     \vertex[right = .75cm of v3](k){$\nu_{\beta}$};
     \diagram*[small]{(i) --[boson](v1),(v3)--[fermion,edge label = $\xi_{i_+}$](v1)--[majorana,edge label = $\xi_{i_-}$](v2),(v2)--[fermion](j),(v3)--[fermion](k),(v2)--[scalar,edge label=$\eta_I$](v3)};
     \end{feynman}   
     \end{tikzpicture}\\
    \begin{tikzpicture}[/tikzfeynman/small]
     \begin{feynman}
     \vertex (i){$Z_D$};
     \vertex[right = 1.cm of i](v1);
     \vertex[right = 1.cm of v1](v4);
     \vertex[above = 0.5cm of v4](v2);
     \vertex[below = 0.5cm of v4](v3);
     \vertex[right = .75cm of v2](j){$\nu_{\alpha}$};
     \vertex[right = .75cm of v3](k){$\nu_{\beta}$};
     \diagram*[small]{(i) --[boson](v1),(v3)--[majorana,edge label = $\xi_{i_+}$](v1)--[anti fermion,edge label = $\xi_{i_+}$](v2),(v2)--[fermion](j),(v3)--[fermion](k),(v2)--[scalar,edge label=$\eta_R$](v3)};
     \end{feynman}   
     \end{tikzpicture} & 
     \begin{tikzpicture}[/tikzfeynman/small]
     \begin{feynman}
     \vertex (i){$Z_D$};
     \vertex[right = 1.cm of i](v1);
     \vertex[right = 1.cm of v1](v4);
     \vertex[above = 0.5cm of v4](v2);
     \vertex[below = 0.5cm of v4](v3);
     \vertex[right = .75cm of v2](j){$\nu_{\alpha}$};
     \vertex[right = .75cm of v3](k){$\nu_{\beta}$};
     \diagram*[small]{(i) --[boson](v1),(v3)--[fermion,edge label = $\xi_{i_+}$](v1)--[majorana,edge label = $\xi_{i_+}$](v2),(v2)--[fermion](j),(v3)--[fermion](k),(v2)--[scalar,edge label=$\eta_R$](v3)};
     \end{feynman}   
     \end{tikzpicture} &
    \begin{tikzpicture}[/tikzfeynman/small]
     \begin{feynman}
     \vertex (i){$Z_D$};
     \vertex[right = 1.cm of i](v1);
     \vertex[right = 1.cm of v1](v4);
     \vertex[above = 0.5cm of v4](v2);
     \vertex[below = 0.5cm of v4](v3);
     \vertex[right = .75cm of v2](j){$\nu_{\alpha}$};
     \vertex[right = .75cm of v3](k){$\nu_{\beta}$};
     \diagram*[small]{(i) --[boson](v1),(v3)--[majorana,edge label = $\xi_{i_+}$](v1)--[anti fermion,edge label = $\xi_{i_+}$](v2),(v2)--[fermion](j),(v3)--[fermion](k),(v2)--[scalar,edge label=$\eta_I$](v3)};
     \end{feynman}   
     \end{tikzpicture} &
     \begin{tikzpicture}[/tikzfeynman/small]
     \begin{feynman}
     \vertex (i){$Z_D$};
     \vertex[right = 1.cm of i](v1);
     \vertex[right = 1.cm of v1](v4);
     \vertex[above = 0.5cm of v4](v2);
     \vertex[below = 0.5cm of v4](v3);
     \vertex[right = .75cm of v2](j){$\nu_{\alpha}$};
     \vertex[right = .75cm of v3](k){$\nu_{\beta}$};
     \diagram*[small]{(i) --[boson](v1),(v3)--[fermion,edge label = $\xi_{i_+}$](v1)--[majorana,edge label = $\xi_{i_+}$](v2),(v2)--[fermion](j),(v3)--[fermion](k),(v2)--[scalar,edge label=$\eta_I$](v3)};
     \end{feynman}   
     \end{tikzpicture}\\
         \begin{tikzpicture}[/tikzfeynman/small]
     \begin{feynman}
     \vertex (i){$Z_D$};
     \vertex[right = 1.cm of i](v1);
     \vertex[right = 1.cm of v1](v4);
     \vertex[above = 0.5cm of v4](v2);
     \vertex[below = 0.5cm of v4](v3);
     \vertex[right = .75cm of v2](j){$\nu_{\alpha}$};
     \vertex[right = .75cm of v3](k){$\nu_{\beta}$};
     \diagram*[small]{(i) --[boson](v1),(v3)--[majorana,edge label = $\xi_{i_-}$](v1)--[anti fermion,edge label = $\xi_{i_-}$](v2),(v2)--[fermion](j),(v3)--[fermion](k),(v2)--[scalar,edge label=$\eta_R$](v3)};
     \end{feynman}   
     \end{tikzpicture} & 
     \begin{tikzpicture}[/tikzfeynman/small]
     \begin{feynman}
     \vertex (i){$Z_D$};
     \vertex[right = 1.cm of i](v1);
     \vertex[right = 1.cm of v1](v4);
     \vertex[above = 0.5cm of v4](v2);
     \vertex[below = 0.5cm of v4](v3);
     \vertex[right = .75cm of v2](j){$\nu_{\alpha}$};
     \vertex[right = .75cm of v3](k){$\nu_{\beta}$};
     \diagram*[small]{(i) --[boson](v1),(v3)--[fermion,edge label = $\xi_{i_-}$](v1)--[majorana,edge label = $\xi_{i_-}$](v2),(v2)--[fermion](j),(v3)--[fermion](k),(v2)--[scalar,edge label=$\eta_R$](v3)};
     \end{feynman}   
     \end{tikzpicture} &
    \begin{tikzpicture}[/tikzfeynman/small]
     \begin{feynman}
     \vertex (i){$Z_D$};
     \vertex[right = 1.cm of i](v1);
     \vertex[right = 1.cm of v1](v4);
     \vertex[above = 0.5cm of v4](v2);
     \vertex[below = 0.5cm of v4](v3);
     \vertex[right = .75cm of v2](j){$\nu_{\alpha}$};
     \vertex[right = .75cm of v3](k){$\nu_{\beta}$};
     \diagram*[small]{(i) --[boson](v1),(v3)--[majorana,edge label = $\xi_{i_-}$](v1)--[anti fermion,edge label = $\xi_{i_-}$](v2),(v2)--[fermion](j),(v3)--[fermion](k),(v2)--[scalar,edge label=$\eta_I$](v3)};
     \end{feynman}   
     \end{tikzpicture} &
     \begin{tikzpicture}[/tikzfeynman/small]
     \begin{feynman}
     \vertex (i){$Z_D$};
     \vertex[right = 1.cm of i](v1);
     \vertex[right = 1.cm of v1](v4);
     \vertex[above = 0.5cm of v4](v2);
     \vertex[below = 0.5cm of v4](v3);
     \vertex[right = .75cm of v2](j){$\nu_{\alpha}$};
     \vertex[right = .75cm of v3](k){$\nu_{\beta}$};
     \diagram*[small]{(i) --[boson](v1),(v3)--[fermion,edge label = $\xi_{i_-}$](v1)--[majorana,edge label = $\xi_{i_-}$](v2),(v2)--[fermion](j),(v3)--[fermion](k),(v2)--[scalar,edge label=$\eta_I$](v3)};
     \end{feynman}   
     \end{tikzpicture}
    \end{tabular}
    \caption{One-loop contributions mediated by scalars to the effective vertex presented in Fig. \ref{fig:eff_ZD}. We adopt two-spinor notation presented in \cite{Dreiner:2008tw}   for drawing the diagrams and calculations.}
    \label{fig:loopf}
\end{figure}

In this section we present the details of the effective $Z_D$ coupling depicted as the bolb shown in Fig. \ref{fig:eff_ZD}.
The Feynman diagrams for the 1-loop contributions to the effective coupling are shown
in Figs. \ref{fig:loopf} and \ref{fig:loops}.
The 1-loop contributions in Fig. \ref{fig:loopf} are mediated by scalars, whereas
those in Fig. \ref{fig:loops} are mediated by fermions.

The contributions from the top 2 rows in Fig. \ref{fig:loopf} lead to the effective Lagrangian given as:
\begin{align}
    \mathcal{L}^a_{eff} &= 2i\nu^\dagger_\alpha\overline{\sigma}^{\mu \nu}\nu^\dagger_{\beta}(p_\alpha-p_\beta)_\nu\sin{2\theta}\left[(M_{i^-}-M_{i^+})(\mathcal{F}_+(m_{\eta_R},M_{i^+},M_{i^-})-\mathcal{F}_+(m_{\eta_I},M_{i^+},M_{i^-}))\nonumber \right. \\
    &+ \left. (M_{i^-}+M_{i^+})(\mathcal{F}_-(m_{\eta_R},M_{i^+},M_{i^-})-\mathcal{F}_-(m_{\eta_I},M_{i^+},M_{i^-}))\right], 
\end{align}
where the mixing angle $\theta$ is given in eq. \ref{eq:mix}, and the contributions coming from the bottom 2 rows in Fig. \ref{fig:loopf} give rise to the effective Lagrangian as follows:
\begin{align}
    \mathcal{L}^b_{eff} &= 2i\nu^\dagger_\alpha\overline{\sigma}^{\mu \nu}\nu^\dagger_{\beta}(p_\alpha-p_\beta)_\nu\cos{2\theta}\left[M_{i^-}(\mathcal{F}_-(m_{\eta_R},M_{i^-},M_{i^-})-\mathcal{F}_-(m_{\eta_I},M_{i^-},M_{i^-})) \nonumber \right. \\
    &- \left. M_{i^+}(\mathcal{F}_-(m_{\eta_R},M_{i^+},M_{i^+})-\mathcal{F}_-(m_{\eta_I},M_{i^+},M_{i^+}))\right].
\end{align}
Similarly, the effective Lagrangian from the contributions shown in Fig. \ref{fig:loops} is given as:
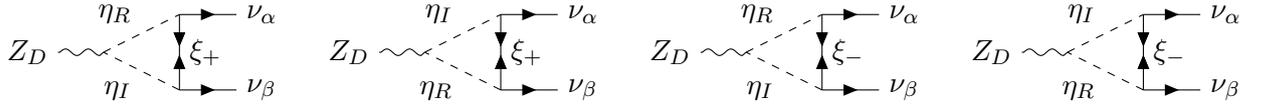
\begin{figure}
    \centering
        \begin{tabular}{lccr}
        \begin{tikzpicture}[/tikzfeynman/small]
     \begin{feynman}
     \vertex (i){$Z_D$};
     \vertex[right = 1.cm of i](v1);
     \vertex[right = 1.cm of v1](v4);
     \vertex[above = 0.5cm of v4](v2);
     \vertex[below = 0.5cm of v4](v3);
     \vertex[right = .75cm of v2](j){$\nu_{\alpha}$};
     \vertex[right = .75cm of v3](k){$\nu_{\beta}$};
     \diagram*[small]{(i) --[boson](v1),(v3)--[scalar,edge label = $\eta_I$](v1)--[scalar,edge label = $\eta_R$](v2),(v2)--[fermion](j),(v3)--[fermion](k),(v2)--[majorana,edge label=$\xi_+$](v3)};
     \end{feynman}   
     \end{tikzpicture} & 
     \begin{tikzpicture}[/tikzfeynman/small]
     \begin{feynman}
     \vertex (i){$Z_D$};
     \vertex[right = 1.cm of i](v1);
     \vertex[right = 1.cm of v1](v4);
     \vertex[above = 0.5cm of v4](v2);
     \vertex[below = 0.5cm of v4](v3);
     \vertex[right = .75cm of v2](j){$\nu_{\alpha}$};
     \vertex[right = .75cm of v3](k){$\nu_{\beta}$};
     \diagram*[small]{(i) --[boson](v1),(v3)--[scalar,edge label = $\eta_R$](v1)--[scalar,edge label = $\eta_I$](v2),(v2)--[fermion](j),(v3)--[fermion](k),(v2)--[majorana,edge label=$\xi_+$](v3)};
     \end{feynman}   
     \end{tikzpicture} &
    \begin{tikzpicture}[/tikzfeynman/small]
     \begin{feynman}
     \vertex (i){$Z_D$};
     \vertex[right = 1.cm of i](v1);
     \vertex[right = 1.cm of v1](v4);
     \vertex[above = 0.5cm of v4](v2);
     \vertex[below = 0.5cm of v4](v3);
     \vertex[right = .75cm of v2](j){$\nu_{\alpha}$};
     \vertex[right = .75cm of v3](k){$\nu_{\beta}$};
     \diagram*[small]{(i) --[boson](v1),(v3)--[scalar,edge label = $\eta_I$](v1)--[scalar,edge label = $\eta_R$](v2),(v2)--[fermion](j),(v3)--[fermion](k),(v2)--[majorana,edge label=$\xi_-$](v3)};
     \end{feynman}   
     \end{tikzpicture} &
     \begin{tikzpicture}[/tikzfeynman/small]
     \begin{feynman}
     \vertex (i){$Z_D$};
     \vertex[right = 1.cm of i](v1);
     \vertex[right = 1.cm of v1](v4);
     \vertex[above = 0.5cm of v4](v2);
     \vertex[below = 0.5cm of v4](v3);
     \vertex[right = .75cm of v2](j){$\nu_{\alpha}$};
     \vertex[right = .75cm of v3](k){$\nu_{\beta}$};
     \diagram*[small]{(i) --[boson](v1),(v3)--[scalar,edge label = $\eta_R$](v1)--[scalar,edge label = $\eta_I$](v2),(v2)--[fermion](j),(v3)--[fermion](k),(v2)--[majorana,edge label=$\xi_-$](v3)};
     \end{feynman}   
     \end{tikzpicture}
     \end{tabular}
    \caption{One-loop contributions mediated by fermions to the effective vertex.}
    \label{fig:loops}
\end{figure}
\begin{align}
    \mathcal{L}^c_{eff} &= -2i\nu^\dagger_\alpha\overline{\sigma}^{\mu \nu}\nu^\dagger_{\beta}(p_\alpha-p_\beta)_\nu(2\cos^2{\theta}\mathcal{F}_+(M_{i^+},m_{\eta_R},m_{\eta_I}) + 2\sin^2{\theta}\mathcal{F}_+(M_{i^-},m_{\eta_R},m_{\eta_I}))
\end{align}
So, combining all the contributions the final form of the effective Lagrangian is given as:
\begin{align}
    \mathcal{L}_{eff} &= \mathcal{L}^a_{eff} + \mathcal{L}^b_{eff} + \mathcal{L}^c_{eff} \nonumber \\
    &= 2i\nu^\dagger_\alpha\overline{\sigma}^{\mu \nu}\nu^\dagger_{\beta}(p_\alpha-p_\beta)_\nu\left[(M_{i^-}-M_{i^+})(\mathcal{F}_+(m_{\eta_R},M_{i^+},M_{i^-})-\mathcal{F}_+(m_{\eta_I},M_{i^+},M_{i^-}))\nonumber \right. \\
    &+ \left. (M_{i^-}+M_{i^+})(\mathcal{F}_-(m_{\eta_R},M_{i^+},M_{i^-})-\mathcal{F}_-(m_{\eta_I},M_{i^+},M_{i^-}))\right. \nonumber \\
    &+ \left. M_{i^-}(\mathcal{F}_-(m_{\eta_R},M_{i^-},M_{i^-})-\mathcal{F}_-(m_{\eta_I},m_{\xi_-},m_{\xi_-})) \nonumber \right. \\
    &- M_{i^+}(\mathcal{F}_-(m_{\eta_R},M_{i^+},M_{i^+})-\mathcal{F}_-(m_{\eta_I},M_{i^+},M_{i^+})) \nonumber \\
    &- \left. 2\cos^2{\theta}\mathcal{F}_+(M_{i^+},m_{\eta_R},m_{\eta_I}) - 2\sin^2{\theta}\mathcal{F}_+(M_{i^-},m_{\eta_R},m_{\eta_I})\right] \nonumber \\
    &= 2i\nu^\dagger_\alpha\overline{\sigma}^{\mu \nu}\nu^\dagger_{\beta}(p_\alpha-p_\beta)_\nu f_i.
    \label{effectiveC}
\end{align}
\section{Details of the loop function $\mathcal{F}_\pm$}
\label{sec:app2}
In order to derive the $\mathcal{F}_\pm$ functions we first start with the well known scalar integral functions 
\begin{align}
    \mathcal{C}_0(p^2_1,p^2_2,p_1.p_2,m^2_1,m^2_2,m^2_3) &= \int \frac{dl^4}{2\pi^4}\frac{1}{(l^2-m^2_1)((l-p_1)^2-m^2_2)((l+p_2)^2-m^2_3)} \nonumber \\
    \mathcal{B}_0(p^2,m^2_1,m^2_2) &= \int \frac{dl^4}{2\pi^4}\frac{1}{(l^2-m^2_1)((l+p)^2-m^2_2)}
\end{align}
where $p^2_1=p^2_2=0$ for outgoing neutrinos in our case and $p1.p_2=s/2$.
We need to calculate the following integral 
\begin{align}
    \int \frac{dl^4}{2\pi^4}\frac{l^\mu}{(l^2-m^2_1)((l-p_1)^2-m^2_2)((l+p_2)^2-m^2_3)} &= p^\mu_1\mathcal{C}_1 + p^\mu_2\mathcal{C}_2
\end{align}
where the scalars $\mathcal{C}_1$ and $\mathcal{C}_2$ are functions of the above scalar integrals. 
To obtain $\mathcal{C}_1$ and $\mathcal{C}_2$, we consider the following integral,
\begin{align}
    \int \frac{dl^4}{2\pi^4}\frac{1}{(l^2-m^2_1)}\left[\frac{1}{(l-p_1)^2-m^2_2} - \frac{1}{(l+p_2)^2-m^2_3}\right] &= \mathcal{B}_0(0,m^2_1,m^2_2) - \mathcal{B}_0(0,m^2_1,m^2_3) \nonumber \\
    \int \frac{dl^4}{2\pi^4}\frac{2l.p_1 + 2l.p_2 + m^2_2 - m^2_3}{(l^2-m^2_1)((l-p_1)^2-m^2_2)((l+p_2)^2-m^2_3)} &= \mathcal{B}_0(0,m^2_1,m^2_2) - \mathcal{B}_0(0,m^2_1,m^2_3)
    \end{align}
    which simplifies to
    \begin{align}
    2p_1.p_2(\mathcal{C}_1 + \mathcal{C}_2) &= \mathcal{B}_0(0,m^2_1,m^2_2) - \mathcal{B}_0(0,m^2_1,m^2_3)+ (m^2_3 - m^2_2)\mathcal{C}_0(0,0,s/2,m^2_1,m^2_2,m^2_3)\nonumber \\
    \mathcal{C}_1 + \mathcal{C}_2 &= \frac{1}{s}\left[\mathcal{B}_0(0,m^2_1,m^2_2) - \mathcal{B}_0(0,m^2_1,m^2_3) + (m^2_3 - m^2_2)\mathcal{C}_0(0,0,s/2,m^2_1,m^2_2,m^2_3) \right] 
\end{align}
Now, we extract $\mathcal{C}_2$ by considering the following integral,
\begin{align}
    \int \frac{dl^4}{2\pi^4}\frac{1}{(l+p_2)^2-m^2_3}\left[\frac{1}{l^2-m^2_1} - \frac{1}{(l-p_1)^2-m^2_2}\right] &= \mathcal{B}_0(0,m^2_1,m^2_3) \nonumber \\
    &- \int \frac{dl^4}{2\pi^4}\frac{1}{((l+p_2)^2-m^2_3)((l-p_1)^2-m^2_2)}.
\end{align}
Making the transformation $l\rightarrow k +p_1$ we get
\begin{align}
    \int \frac{dl^4}{2\pi^4}\frac{1}{(l+p_2)^2-m^2_3}\left[\frac{1}{l^2-m^2_1} - \frac{1}{(l-p_1)^2-m^2_2}\right] &= \mathcal{B}_0(0,m^2_1,m^2_3) \nonumber \\
    &- \int \frac{dk^4}{2\pi^4}\frac{1}{((k + p_1 +p_2)^2-m^2_3)(k^2-m^2_2)} \nonumber \\
    \int \frac{dl^4}{2\pi^4}\frac{m^2_1 - m^2_2 - 2l.p_1}{(l^2-m^2_1)((l-p_1)^2-m^2_2)((l+p_2)^2-m^2_3)}&= \mathcal{B}_0(0,m^2_1,m^2_3) - \mathcal{B}_0(s,m^2_2,m^2_3) 
\end{align}
\begin{align}
    -2p_1.p_2\mathcal{C}_2 &= \mathcal{B}_0(0,m^2_1,m^2_3) - \mathcal{B}_0(s,m^2_2,m^2_3) + (m^2_1 - m^2_2)\mathcal{C}_0(0,0,s/2,m^2_1,m^2_2,m^2_3) \nonumber \\
    \mathcal{C}_2 &= \frac{1}{2p_1.p_2}\left[\mathcal{B}_0(s,m^2_2,m^2_3) - \mathcal{B}_0(0,m^2_1,m^2_3) + (m^2_2 - m^2_1)\mathcal{C}_0(0,0,s/2,m^2_1,m^2_2,m^2_3)\right] \nonumber \\
    \mathcal{C}_2 &= \frac{1}{s}\left[\mathcal{B}_0(s,m^2_2,m^2_3) - \mathcal{B}_0(0,m^2_1,m^2_3) + (m^2_2 - m^2_1)\mathcal{C}_0(0,0,s/2,m^2_1,m^2_2,m^2_3)\right]
\end{align}
From the above equation, we get  
\begin{align}
    \mathcal{C}_1 &= \frac{1}{s}\left[\mathcal{B}_0(s,m^2_1,m^2_2) - \mathcal{B}_0(s,m^2_2,m^2_3) + (m^2_3 - m^2_1)\mathcal{C}_0(0,0,s/2,m^2_1,m^2_2,m^2_3)\right]
\end{align}
and from the scalar $\mathcal{C}_1$ and $\mathcal{C}_2$ we construct the $\mathcal{F}$'s as given below
\begin{align}
    \mathcal{F}_+(m_1,m_2,m_3) &= \mathcal{C}_1 + \mathcal{C}_2 \nonumber \\
    \mathcal{F}_+(m_1,m_2,m_3) &= \frac{1}{s}\left[\mathcal{B}_0(0,m^2_1,m^2_2) - \mathcal{B}_0(0,m^2_1,m^2_3) + (m^2_3 - m^2_2)\mathcal{C}_0(0,0,s/2,m^2_1,m^2_2,m^2_3) \right] \nonumber \\
    \mathcal{F}_-(m_1,m_2,m_3) &= \mathcal{C}_1 - \mathcal{C}_2 \nonumber \\
    \mathcal{F}_-(m_1,m_2,m_3)&= \frac{1}{s}\left[\mathcal{B}_0(0,m^2_1,m^2_2) + \mathcal{B}_0(0,m^2_1,m^2_3) - 2\mathcal{B}_0(s,m^2_2,m^2_3) \right. \nonumber \\
    &+ \left.(m^2_2 + m^2_3 - 2m^2_1)\mathcal{C}_0(0,0,s/2,m^2_1,m^2_2,m^2_3)\right]
\end{align}
In our work we have used the {\tt Package-X}\cite{Patel:2015tea,Patel:2016fam} in order to calculate the scalar integrals $\mathcal{C}_0$ and $\mathcal{B}_0$ mentioned above.
\acknowledgments
The work of SKK and AD was support in part by  the National Research Foundation (NRF) grant  NRF-2019R1A2C1088953.
\bibliographystyle{JHEP}
\bibliography{ref}
\end{document}